\pdfoutput=1
\documentclass[print]{iucr}
    \journalcode{S}

\usepackage{xcolor}
\usepackage{graphicx}
\begin{document}
\title{Speckle correlation as a monitor of X-ray free electron laser induced crystal lattice deformation}
%\title{A speckle correlation scheme for monitoring X-ray free electron laser induced lattice disorder}
%\title{Observation of X-ray free electron laser-induced disorder in a single crystal using two-time speckle correlation analysis}

\author[a]{Rajan}{Plumley}
\author[a]{Yanwen}{Sun}
\author[a,b]{Samuel}{Teitelbaum}
\author[a]{Sanghoon}{Song}
\author[a]{Takahiro}{Sato}
\author[a]{Matthieu}{Chollet}
\author[a]{Nan}{Wang}
\author[a]{Aymeric}{Robert}
\author[a]{Paul}{Fuoss}
\author[c]{Mark}{Sutton}
\cauthor[a]{Diling}{Zhu}{dlzhu@slac.stanford.edu}

\aff[a]{Linac Coherent Light Source, SLAC National Accelerator Laboratory, 2575 Sand Hill Road, Menlo Park, CA 94025, \country{USA}}
\aff[b]{PULSE Institute, SLAC National Accelerator Laboratory, 2575 Sand Hill Road, Menlo Park, CA 94025, \country{USA}}
\aff[c]{Physics Department, McGill University, 845 Sherbrooke St W, Montr\'eal, Quebec H3A 0G4, \country{Canada}}

\begin{synopsis}
In this work we investigate the FEL X-ray beam-induced lattice disorder by measuring shot-to-shot evolution of near-Bragg coherent diffuse scattering from single-crystal germanium. We show that X-ray photon correlation analysis of sequential speckle measurements can be used to monitor the nature and extent of the lattice displacement rearrangement.
\end{synopsis}

\begin{abstract}
X-ray free electron lasers (X-FELs) present new opportunities to study ultrafast lattice dynamics in complex materials. While the unprecedented source brilliance enables high fidelity measurement of structural dynamics, it also raises experimental challenges related to the understanding and control of beam-induced irreversible structural changes in samples that can ultimately impact the interpretation of experimental results. This is also important for designing reliable high performance X-ray optical components. In this work, we investigate X-FEL beam-induced lattice alterations by measuring the shot-to-shot evolution of near-Bragg coherent scattering from a single crystalline germanium sample. We show that X-ray photon correlation analysis of sequential speckle patterns measurements can be used to monitor the nature and extent of lattice rearrangements. Abrupt, irreversible changes are observed following intermittent high-fluence monochromatic X-ray pulses, thus revealing the existence of a threshold response to X-FEL pulse intensity.
\\
\\
\end{abstract}

\section{Introduction}

% motivate by 'cumulative damage' the beam can incur to material science samples, optics ...
From studies of condensed matter to biological systems, the exceptionally high brightness of X-FEL pulses has enabled obtaining information at unprecedented small and fast spatial and temporal regimes \cite{Bostedt2016rmp}. However, that same quality also exacerbates the risk of incurring damages to the sample under investigation by the probing X-ray radiation itself. This technical challenge has been a recurring concern for X-ray methods since their beginnings in the late nineteenth century~\cite{Slater1955,henderson1995}. FELs have for example demonstrated that intense X-ray pulses focused down to very small sizes can destroy materials in a single-shot and can even affect X-ray optical components~\cite{HauRiege2008APL, HauRiege2010,Koyama:13}. This reinforces that it is essential to characterize and also understand FEL X-ray beam-induced damage process.

% X-FELs are billions of times brighter than previously available X-ray sources, and it has been shown that single X-FEL pulses can destroy any material when focused -- including the very optics used to manipulate and deliver them to the sample. Thus it is clear that the characterization and understanding of X-ray FEL-induced damage processes is of great importance.

While the potential risk of incurring radiation damage is a concern for nearly all X-FEL experiments, its severity strongly varies on various parameters such as the modes of operation, the material of interest, and the X-ray measurement technique being employed. The seminal work by Neutze \emph{et al.} described the key experimental concept of obtaining ``diffraction-before-destruction" for the structure determination of macro biomolecules, in which the X-ray diffraction out-runs the Coulomb explosion and ensuing disintegration of the molecule thanks to the femtosecond pulse duration and the `instantaneous' nature of the X-ray scattering process~\cite{Neutze2000}. This concept has also been extended to crystalline samples, and is further strengthened by the notion of ``self-termination of diffraction" which addresses the faster than expected structural alteration by the extremely intense FEL pulses~\cite{boutet2012high,barty2012self}. This also enables damage-free structural determination of biomolecules such as membrane proteins and metalloprotein that are otherwise too sensitive to cumulative damage effects induced by the effectively continuous radiation delivery of third-generation synchrotron sources~\cite{Kern2014}.

In contrast to common conceptions, not all FEL experiments operate in the diffract-before-destroy regime. For example, ultrafast diffraction experiments probing excited structural dynamics in condensed matter typically use stroboscopic measurements where the sample receives repetitive localized exposures from X-ray pulses over hours~\cite{Trigo2013,Clark2013,Gerber2017}. The instantaneous radiation dose must be limited to a low enough level to reduce the risk of cumulative irreversible modification to the sample that could change the dynamics under investigation. For X-ray photon correlation spectroscopy (XPCS) and in particular split-pulse XPCS, the probe-probe nature of the method imposes even more stringent requirement on the intensity of single pulse to minimize its impact on the dynamics being investigated~\cite{Lehmkuhler2018,Moller2019kk}. Carnis \emph{et al.} have used X-FEL radiation to examine the relaxation dynamics of gold nanoparticles while also characterizing the eventual occurrence of sample damage. They demonstrated the feasibility of XPCS studies of soft matter materials at X-FEL beamlines with the caveat that experimental boundaries must be imposed to limit sample degradations~\cite{Carnis:14}. To date, experimental efforts have mostly relied on  empirical `trial and error' approaches. Systematic understanding of sample damage mechanisms has been scarcely documented.

Another challenge unique to FEL sources is the intrinsic intensity fluctuation of SASE-generated X-ray pulses~\cite{Bonifacio1994}. Not only does this further complicate the optimization of XPCS measurements, but if the beamline is operating in a configuration that experiences large shot-to-shot incident intensity variations, the sudden occurrence of a high intensity pulse could damage the sample or even beamline optics. For example, Koyama and coworkers at the SPring-8 angstrom compact free-electron laser facility (SACLA) have observed the single-shot ablation of thin films and substrates commonly used for X-ray optics applications by monochromatic hard X-rays, observing a threshold response and imprint size directly correlated with to the X-ray pulse intensity~\cite{Koyama:13}. 

In fact, the so-called `imprint' technique has been widely used as a way to quantify beam profiles of pulsed laser sources~\cite{Liu:82,HauRiege2010,Koyama:13}. While useful, the method relied on the systematic examination of the dose-array exposed samples at a later time using high resolution optical or electron microscopy. With the commissioning of new X-FEL facilities with even higher brilliance and repetition rates underway, new techniques for monitoring the degradation of materials during the course of experiments will be crucial in identifying optimal measurement conditions. Hence, an on-the-fly, pulse-resolved method compatible with typical scattering experiment setups such that any occurrence of sample alteration can be event-correlated would be desired.

Over the past decades, XPCS studies have demonstrated that the two-time correlation function (TTCF) can be used to study subtle changes in the material structures and dynamics on small length scales with high sensitivity~\cite{Sanborn:11,Evenson:15}. In this work we employ this method to investigate the creation and evolution of lattice disorder in a single-crystalline sample via TTCF analysis of sequential X-ray speckle correlation measurements. We also show that by comparing  FEL X-ray pulse intensities to their resultant lattice rearrangement effects (i.e., as manifested in the decorrelation of subsequent speckle patterns) can be used to determine the threshold for permanent lattice disorder rearrangement. 

\section{Experimental setup and observations}  

\begin{figure}
    \includegraphics[width=\linewidth]{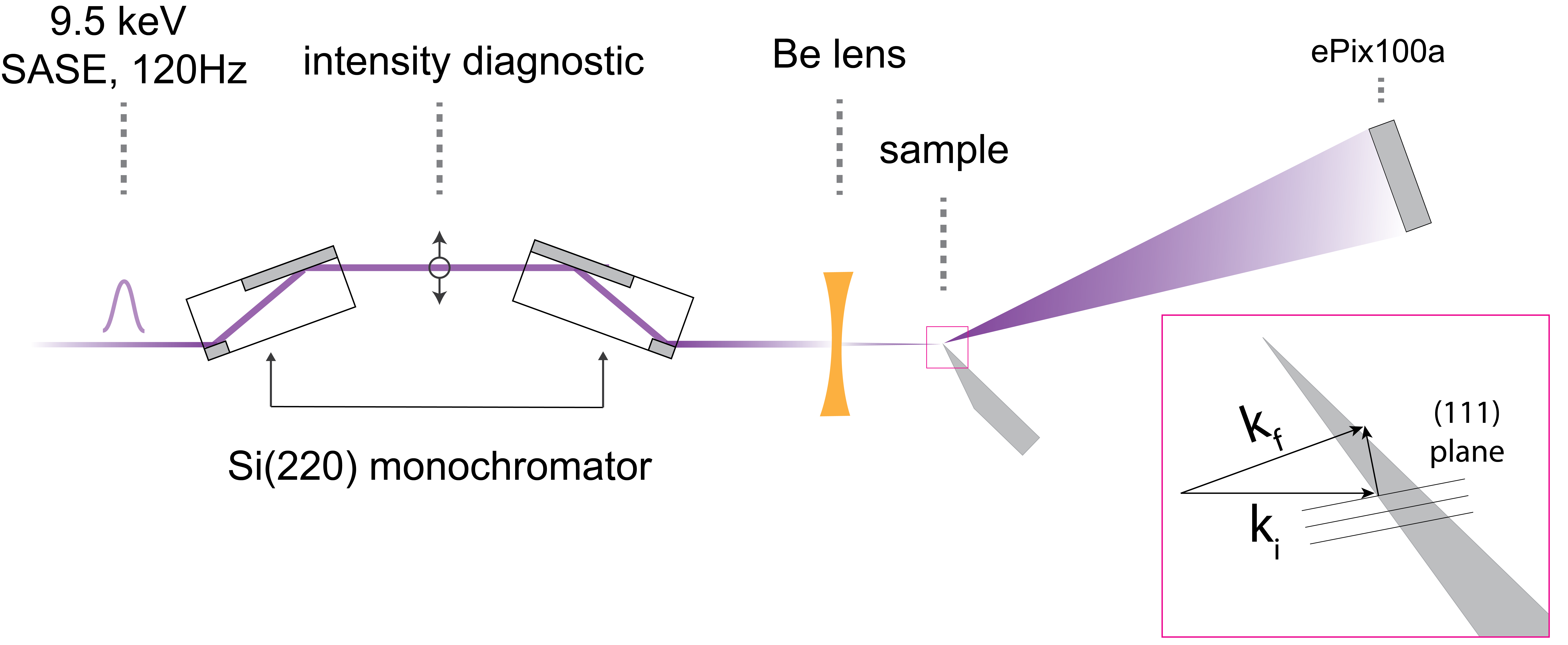}
    \caption{Schematics of the experimental setup. It consists of a 4-bounce Si(220) monochromator using a pair of channelcut crystals, slits for defining the beam trajectory, beryllium CRLs for focusing, and a diffractometer for orienting the sample. The scattering was collected with an ePix100a detector at a distance of ~5.5 meter from the sample. The inset shows the geometry of the crystal `wedge' and the orientation of the Ge(111) lattice plane, as well as the reciprocal space scattering geometry.}
    \label{fig:schematics}
\end{figure}

The experiment was performed at the X-ray Correlation Spectroscopy (XCS) instrument at the Linac Coherent Light Source~\cite{alonso2015x}. As illustrated in Figure \ref{fig:schematics}, an X-ray beam with a photon energy of 9.5~keV~was monochromatized using a dual-channelcut four-bounce Si(220) monochromator (i.e., the fixed-delay branch of the compact split-delay system~\cite{Sun:19} was used) down to a bandwidth of 0.5~eV. The X-ray pulse duration was estimated to be approximately 50~fs. Pulse intensities were monitored in-situ by a photodiode measuring scattering from a Kapton foil between the two Si(220) channelcuts. Beryllium compound refractive lenses (CRLs) with a focal length of 0.9 m were used to focus the beam to approximately 2 micron at the sample location. The germanium single crystal was polished into a 10 degree wedge shape and mounted on a compact goniometer in air at room temperature. The wedge shape allows to adjust the sample thickness with a simple translation; thus enabling the adjustment of the balance between the signal level, speckle size, and speckle contrast. After locating the (111) Bragg peak, the sample was rotated by $1^\circ$ from the peak position around the vertical axis to avoid detector saturation. The detector samples a slice of the reciprocal space near the Ge(111) Brillouin zone center. Downstream of the sample, the scattered X-rays traveled through a vacuum flight path before being recorded by an ePix100a detector at a distance of 5.5~m~\cite{carini2016epix100}. The detector pixel size was $50 \times 50$ microns. Its position was centered near the peak of the initial thermal diffuse scattering near the Ge (111) Bragg reflection direction. \\

%figure recommendation

%%%%%%%%%%%%%%%%%%%%%%%%%%%%%%%%%%%%%%%%%%%%%%%%%%%%%%%%%%%%%%%%%%%%

\begin{figure}
    \includegraphics[width=\linewidth]{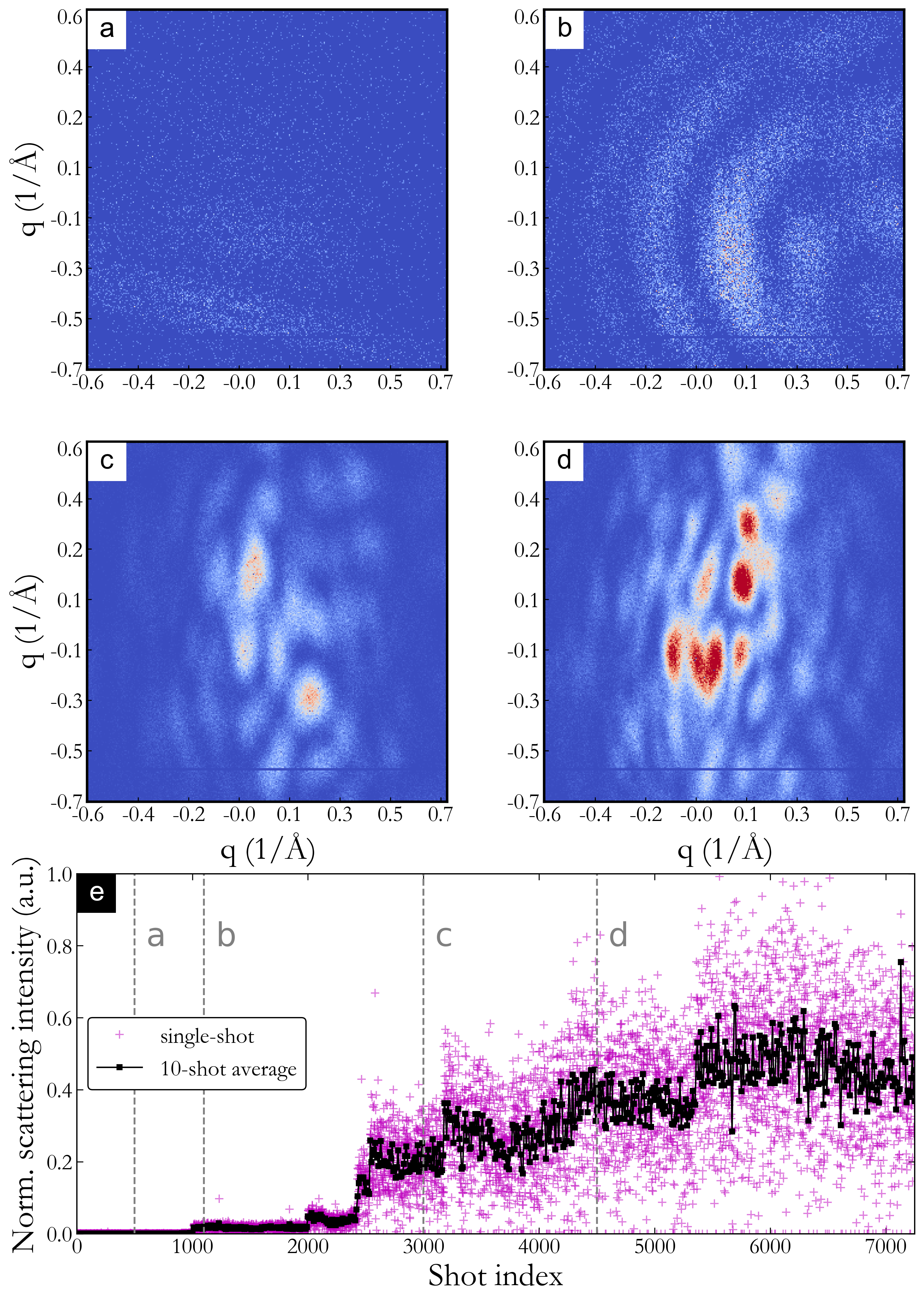}
    \caption{(a-d) Time-averaged near-Bragg scattering speckle patterns showing the evolution of speckles. (e) Single-shot (pink) and 10 shot integrated (black) speckle intensity as a function of total number of X-FEL pulse exposures, showing intensity increase in successive abrupt steps.}    
    \label{fig:speckle_evolution}
\end{figure}

For perfect crystals at finite temperatures, near-Bragg intensities in the form of thermal diffuse scattering (TDS) are a manifestation of dynamic atomic displacements from the ideal crystal lattice positions~\cite{warren1990x}. As larger amplitude static disorder is introduced into the crystal structure, additional diffuse scattering contributes to intensity increases near the zone center. When illuminated with coherent X-rays, these diffuse scattering contributions will appear as X-ray speckles~\cite{Sutton:91}. The speckle pattern is a reflection of the arrangement of the disorder within the scattering volume. The measurements of the evolution of the speckle pattern thus carry information on how the disorder is changing from shot to shot. Figure 2 shows a typical sequence of near-Bragg coherent scattering patterns. Starting from a single crystal in its pristine condition, we observe a weak but relatively uniform distribution of TDS on average as shown in Figure~\ref{fig:speckle_evolution}(a). As the FEL irradiation continues, we observed step-wise increase in the total diffuse scattering intensity. We also observe clearly that the diffuse scattering form `temporarily static' speckle patterns as shown in Figure 2 (b-d). The speckle patterns remained similar for typically a few tens of pulses and then rearranged significantly as seen in the difference between Figure 2 (c) and (d).

The step-wise sudden changes of the speckle pattern are triggered by so-called `hot shots', as monochromatization of the SASE X-FEL beam presents high pulse intensity fluctuations~\cite{zhu2014performance}. On the other hand, we did not observe significant changes in speckle sizes nor a build up of Debye-Scherrer ring, which suggests that the structural modification to the single crystal sample was still relatively subtle. This will be further discussed in a subsequent question.

%%%%%%%%%%%%%%%%%%%%%%%%%%%%%%%%%%%%%%%%%%%%%%%%%%%%%%%%%%%%%%%%%
%\section{Two-time correlation Analysis}

To evaluate the sequence of scattering growth and speckle evolution, we employ the two-time correlation analysis method. The two-time correlation function is an extended formulation of the standard temporal correlation function $g^{(2)}(q, t)$ first used in dynamic light scattering experiments. The TTCF, which measures the statistical similarity between intensity measurements at any two points in time~\cite{Madsen:10}, can be calculated as:
\begin{equation}\label{eq:1}
    g^{(2)}(q,t_1,t_2) = \frac{\langle I(q,t_1) I(q,t_2) \rangle_{\phi}}{\langle I(q,t_1) \rangle_{\phi} \langle I(q,t_2) \rangle_{\phi}},
\end{equation}
where $I(q,t)$ is the pixel intensity at the momentum transfer $q$ and time $t$. $\phi$ in the subscript of the bracket indicates averaging over the pixels where correlation shows negligible variation. It may be convenient to think about $g^{(2)}(q,t_1, t_2)$ as a moving evaluation of $g^{(2)}(q,t)$ such that $t_0$ is always along the $t_1 = t_2$ diagonal.  In these terms, the autocorrelation lag time is $t = t_2 - t_1$. The time-boundary cases of the standard $g^{(2)}(q,t)$ extend to the two-time formulation as follows,
\begin{equation}\label{eq:2}
    g^{(2)}(q,t_1,t_2) = \left\{ \begin{array}{ll} 1+\beta, \,\quad t_1 = t_2 \\ 1, \quad \quad \quad t_2-t_1 \rightarrow \infty \end{array} \right.
\end{equation}
where $\beta$ is the optical contrast~\cite{Madsen:10}. Note that the time unit in the context here are `pulses' rather than the real time as we are interested in the beam induced pulse-to-pulse changes. From the coherent scattering patterns, we selected a $250 \times 250$ pixel region of interest (ROI) from the ePix100a detector images centered near the peak of the thermal diffuse scattering and covering a reduced momentum transfer range  $q_{\phi} \rightarrow \left[ -0.6, 0.4 \right] \mathrm{\AA^{-1}}$. We define reduced momentum transfer with respect to TDS center of mass. While our TTCF ranges over 3200 sequential images, we only present the TTCF up to a lag time of $t = 800$, as speckle decorrelations usually occur within the first few hundred shot increments and the total number of calculations to complete scales with the square of the number of images used. 
\\

\begin{figure}
    \begin{center}
    \includegraphics[width=\linewidth]{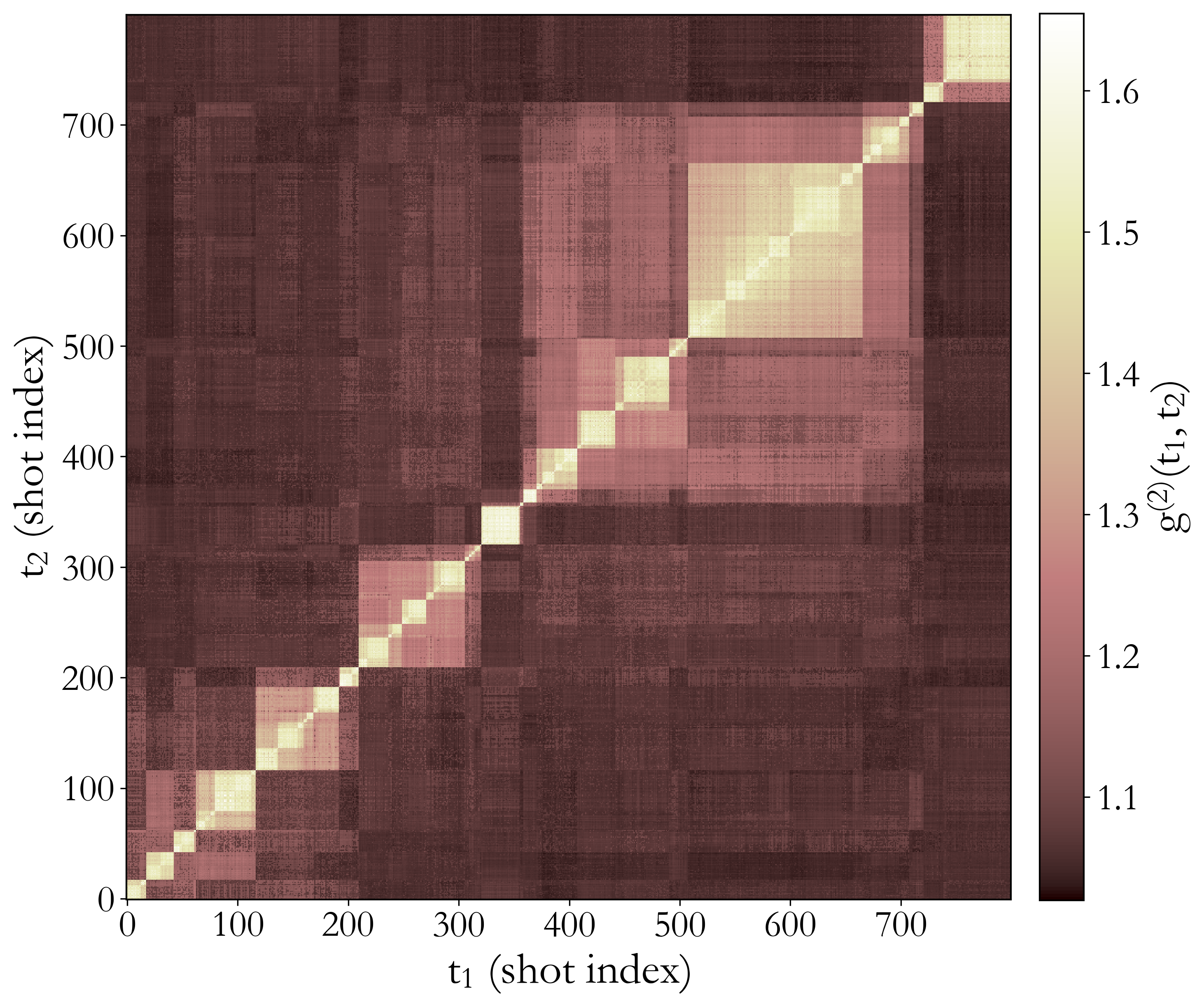}
    \caption{$800 \times 800$ crop of the two-time speckle correlation. Autocorrelation amplitudes lie along the $t_1 = t_2$ diagonal and lag times can be scanned across vertical and horizontal directions.}
    \label{fig:ttcf}
    \end{center}
\end{figure}
Single-pulse exposures were used to calculate the TTCF so that any shot-to-shot evolution of the speckle pattern would not be time-averaged away. This required a high enough scattering intensity to enable sensitivity in speckle visibility measurement for each X-ray pulse~\cite{Falus:06}. The data presented are collected from a slightly thicker part of the sample ($\sim 20$ microns) with an X-ray pulse energy distribution that triggers relatively frequent speckle rearrangement during the experiment (every few seconds). We have also excluded scattering patterns from very low intensity pulses in our analysis as they were too weak to provide an accurate correlation measurement. Pulses at those intensities appeared to be below the threshold that would induce any observable speckle changes based on pulse-averaged measurements with attenuated beams. All speckle patterns were first thresholded at the detector pixel ADU value equivalent to half a photon in order to remove the readout noise of the dark region. Gaussian smoothing is then used to balance the intensity variation across the $q_\phi$ region of interest prior to the calculation of the correlation~\cite{Fluerasu:05}. In total, 3200 sequential speckle pattern images from the ePix100a detector were used. Figure 3 shows an $800 \times 800$ section of the full two-time correlation function matrix. 
\begin{figure}
    \begin{center}
    \includegraphics[width=0.875\linewidth]{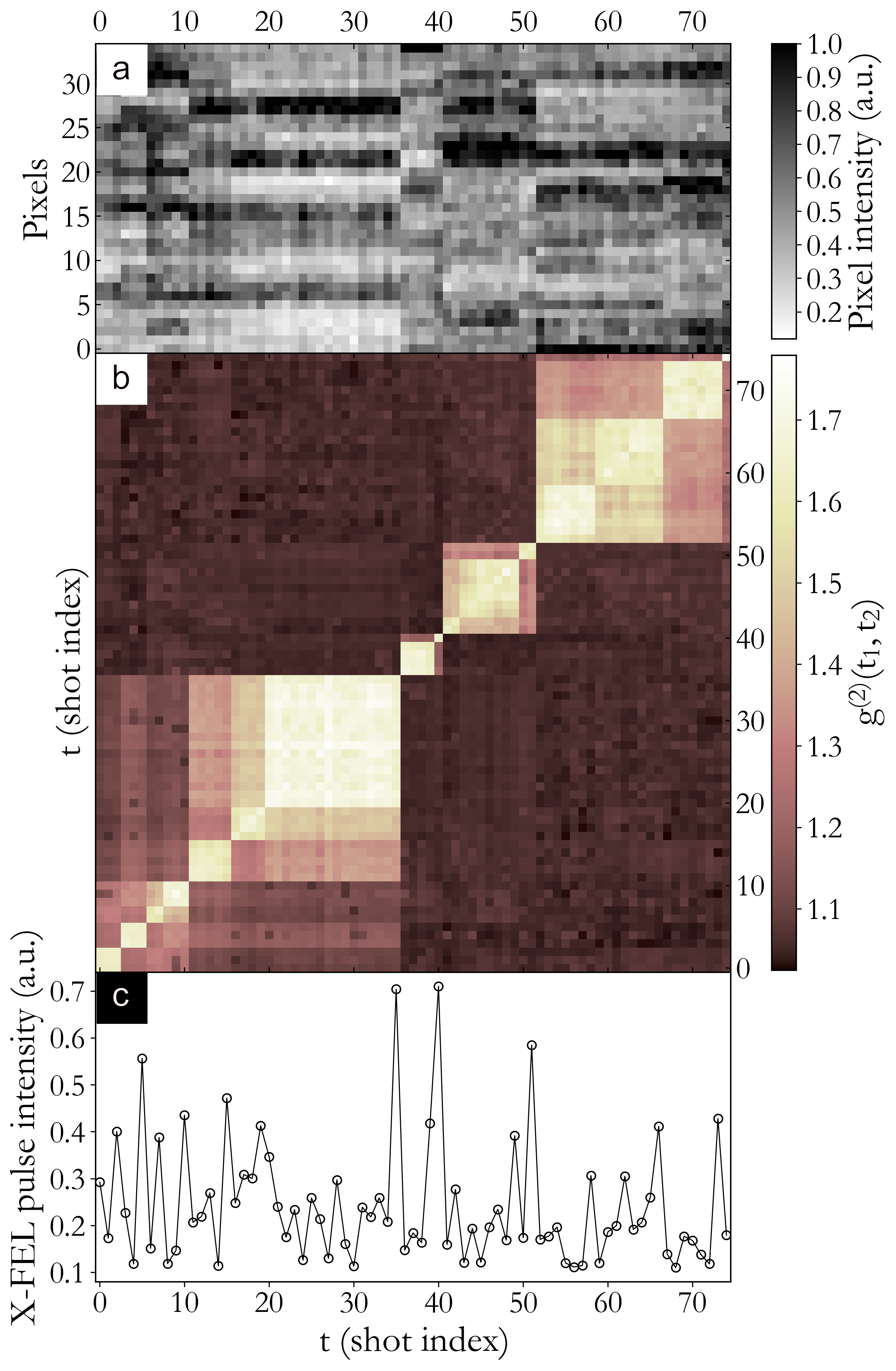}
    \label{fig:sbs}
    \caption{(a) A typical line-out of the speckle pattern as a function of number of shots. (b) a $75 \times 75$ ROI of the TTCF. (c) shot-by-shot X-ray pulse intensity measurements. One sees that the high intensity spikes aligns well with TTCF edges indicates conjunction between high-intensity laser pulses and abrupt speckle decorrelation on the next exposure event.}
    \end{center}
\end{figure}
% figure 4
While two-time correlation analysis is predominantly used to study dynamics in non-equilibrium systems~\cite{Evenson:15, Luttich:18, Madsen:10}, our aim is to examine the abrupt changes of the speckle patterns resulting from stochastic variations in the X-FEL probe pulse intensity. Quick inspection of the TTCF in Figure 3 reveals a sequence of square tiles arranged along the autocorrelation diagonal. Tiles in the TTCF correspond to time-durations of minimal variation in the speckle intensity configuration, while their outward-facing edges characterize decorrelation within the time span between FEL exposure events. This signifies a rapid transformation of the atomic lattice structure similar to the two-time observations of martensitic transformations in cobalt by Sanborn and others~\cite{Sanborn:11}. This behavior is examined more closely in Figure 4. Figure 4 (a) shows an example of speckle intensity line-out as a function of pulses, where abrupt transitions can be directly visualized. In the corresponding TTCF map shown in Figure 4 (b), we can see that the transitions of various degree maps very well to the touching corners of the square tiles of similar correlation values. All the abrupt transitions are also well matched to the shot-by-shot FEL pulse intensity measurement plotted on an aligned horizontal axis in Figure 4(c). Clearly, the extent of `decorrelation' of the speckle pattern from the previous pattern is strongly related to the pulse energy of the previous pulse. To analyze this relationship more quantitatively, we define the speckle reconfiguration amplitude $\delta(t)$, that has the expression 
\begin{equation}\label{eq:3}
    \delta(t) = g^{(2)}(t,t) - g^{(2)}(t,t - 1),
\end{equation}
which is a measure of the change in $g^{(2)}$ amplitude at each time-step.  Because a speckle pattern contains information that reflects the disordered structure producing it~\cite{Sutton:91}, we can use $\delta(t)$ as a shot-by-shot monitor of the extent of lattice rearrangement that occurs in the irradiated portion of the sample. Equation (\ref{eq:3}) is evaluated along the $t = t_1 = t_2$ diagonal for the full range of TTCF data. We use the event-index $t$ to gather pulse intensities $I_0(t-1)$ from the previous FEL shots. A plot of the speckle transition amplitude v.s. X-FEL pulse intensity is shown in Figure 5. 

\begin{figure}
    \begin{center}
    \includegraphics[width=0.9\linewidth]{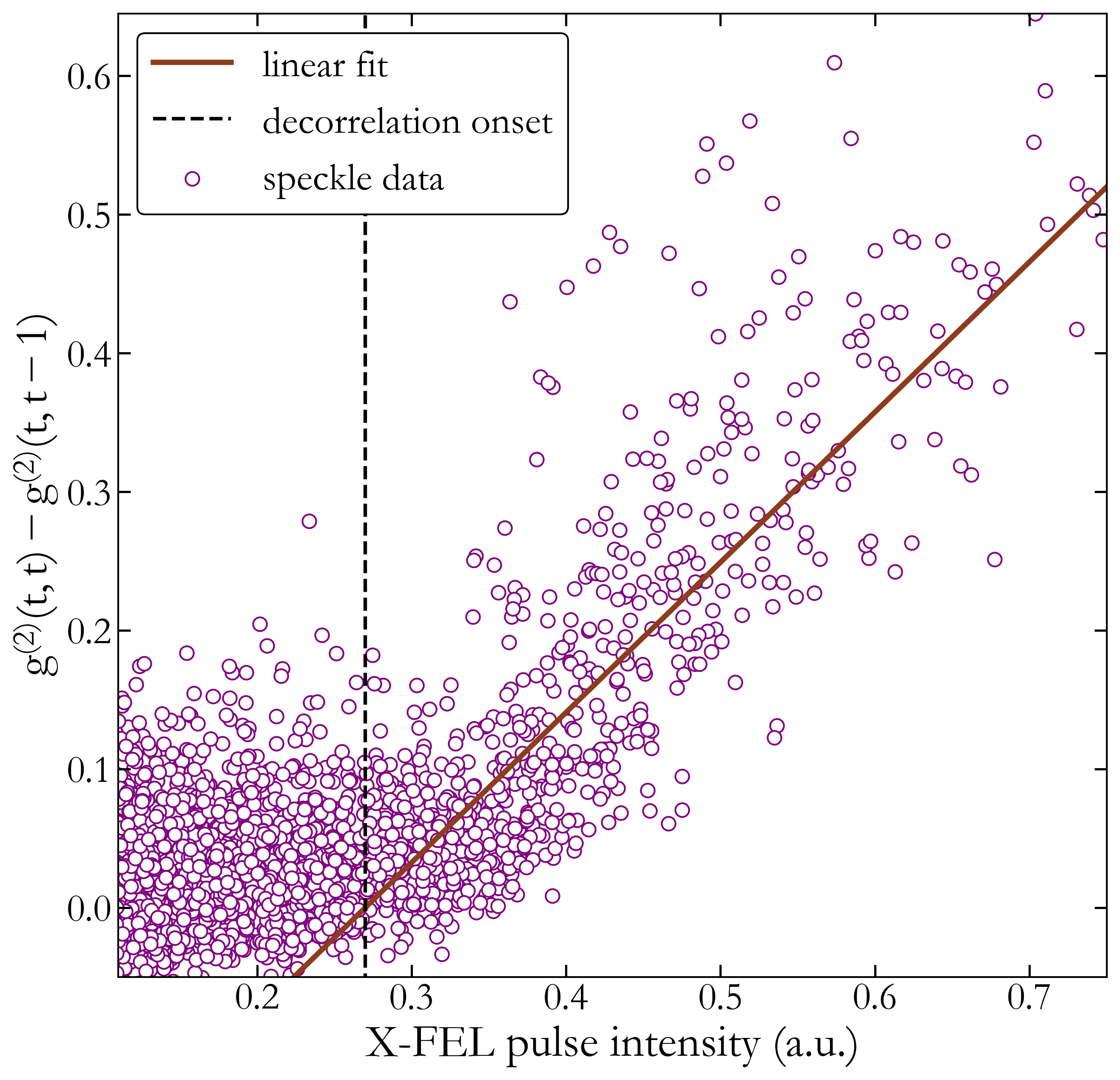}
    \caption{Speckle reconfiguration amplitudes $\delta$ evaluated over 3200 measurements  v.s. X-FEL pulse intensities of their previous shot index. A linear fit to determine the onset-intensity to drive speckle change is plotted (solid line), with the threshold indicated by the dotted line.}
    \label{fig:hockeystick}
    \end{center}
\end{figure}

Figure~\ref{fig:hockeystick} shows a threshold behavior of the speckle reconfiguration amplitude as a function of the X-FEL pulse intensity. While the energy deposition from the X-ray pulse is proportional to the measured pulse energy, there appears to be a minimum pulse energy required induce observable structural changes in the sample that led to the rearrangement of the speckle pattern.  In the plot we show a least-squares linear fit to the data (with a fitting range for X-FEL pulse intensity larger than 0.30 a.u.) can be used to estimate the damaging threshold intensity of 0.27 a.u. However, while this in principle can be a rather sensitive probe to the structural damage threshold, we refrain from the discussion of the absolute threshold as it requires accurate measurement of the beam size, profile, as well as pulse energy which unfortunately was inadequate in this experiment.
\section{Discussion \& Conclusion}

The threshold response of $\delta(t)$ favors the picture that the FEL beam induces a transient thermal gradient within the beam footprint and the center part of the irradiated sample reached a crystalline structural instability close to the melting point of the sample. The best estimate of the absolute pulse energy on the sample also put the threshold dose to within a factor of 2 of the melting dose of Ge at approximately 0.2 eV/atom. The absence of the formation or even a precursor of Debye-Scherrer scattering ring indicates that the resolidification following the likely-melting of the sample did not result in an isotropic polycrystaline structure. It is also highly unlikely that the melted portion of the sample would enter an amorphous phase as that would lead to observation of reduced speckle sizes (as the effective volume of the sample contributing to the preferential scattering towards the $q$-region of interest will be larger). We thus hypothesize that the cooling rate must be sufficiently low such that a partially melted sample volume would have enough time to adiabatically recrystallize following the general lattice orientation of the surrounding cooler part of the crystal. This hypothesis is also supported by the lack of surface ablation features, damage craters, or other commonly-found damage features as carefully examined with both optical and scanning electron microscopes.

%%%%%%%%

In conclusion,  we have demonstrated a coherence-based, pulse-resolved technique for investigating irreversible FEL-induced crystal structural alterations. Two-time correlation analysis of near-Bragg speckle measurements was shown to be a sensitive probe for monitoring the evolution of atomic structural disorder in a single crystal, which will be useful for further studies. We have also shown that crystalline structural damage thresholds relevant to specific reciprocal space regions can be determined by quantitative measurement of the decorrelation between successive speckle patterns and its correlation with the incident X-ray pulse intensities. This method is simple and can be made compatible with many hard X-ray scattering instruments at FEL facilities. With sufficient angular resolution on the detection side, it offers a much more sensitive monitor for the sample condition, as compared to the measurement of the average scattering. 

%We also note that with reduced X-ray flux and a rocking curve measurement, a 3D reconstruction of the crystal lattice displacement field can potentially be obtained with nanometer resolution via Bragg coherent diffractive imaging technique\cite{Clark2013}.

%\cite{red}{A thought for now: this can be a pretty good 'condensed matter sample damage threshold screening process' ... although some of the interesting samples itself probably would already have sufficient disorder to mess things up. Also, this can be too useful for XPCS samples where natural dynamics would already give shot to shot speckle fluctuations. ... OK, so basically it can be a important 'sample health monitor' for single crystalline samples.}

Finally, the observation of intermittent rearrangement of the speckle patterns, and the correlation analysis protocol used to pin point the transitions, are relevant for data-reduction algorithms applicable to capturing rare events~\cite{engels1999single, BESUltrafast}, which is a new opportunity enabled by the upcoming MHz X-FEL sources such as LCLS-II and LCLS-II-HE. We anticipate implementing TTCF-like analysis at the hardware level on the detector to down-select the transitional image arrays and thus play a key role in taking full advantage of high-brightness high-repetition X-FEL sources~\cite{islambek2019fpga}.

%As the repetition rate of X-FEL sources increases in the coming years, data reduction techniques will need to be implemented to reduce the total data payload from beamline experiments \cite{thayer2016data}.  This method (in particular an implementation of equation (\ref{eq:3})) opens the doorway for fast on-the-fly self-similarity examination of the sample as measurements are acquired by the detector electronics, which can be used as a trigger for tagging "interesting" events to be saved in the data acquisition pipeline. 

%% a bit more far fetched --  Link this to 'triggered' rare events which is a primary science case being considered for LCLS-II HE. Reference to the BES document.

%% Ah, then potentially a MHz multi frame detector doing 2-time image correlation frame by frame and only save the 'frames with changes' can be great ... ... need to know what to look though.

\section{Acknowledgments}

RP acknowledges the computing support by Omar Qiujano. Use of the Linac Coherent Light Source (LCLS), SLAC National Accelerator Laboratory, is supported by the U.S. Department of Energy, Office of Science, Office of Basic Energy Sciences under Contract No. DE-AC02-76SF00515.

\bibliography{iucr}
\bibliographystyle{iucr}

\end{document}